\documentstyle[12pt,aasms4]{article}

\begin{document}

\title{Spectropolarimetry of FIRST BAL QSOs\altaffilmark{1}}
\author{M. S. Brotherton, Hien D. Tran, Wil van Breugel}
\affil{Institute of Geophysics and Planetary Physics, Lawrence Livermore National Laboratory, 7000 East Avenue, P.O. Box 808, L413, Livermore, CA 94550; mbrother@igpp.llnl.gov, htran@igpp.llnl.gov, wil@igpp.llnl.gov}

\author{Arjun Dey}
\affil{KPNO/NOAO\altaffilmark{2}, 950 N. Cherry Avenue, P. O. Box 26732, Tuscon,
AZ 85726; dey@noao.edu}

\author{Robert Antonucci}
\affil{Physics Department, University of California at Santa Barbara, Santa Barbara, CA 93106; ski@ginger.physics.ucsb.edu}

\altaffiltext{1}{Based on observations at the W. M. Keck Observatory.}
\altaffiltext{2}{The National Optical Astronomy Observatories are operated by
the Association of Universities for Research in Astronomy under cooperative
agreement with the National Science Foundation.}

\begin{abstract}
We present Keck spectropolarimetry of two rare low-ionization
broad absorption line (BAL) QSOs, FIRST J084044.5+363328
and FIRST J155633.8+351758, that also exhibit narrow absorption lines
from metastable excited levels of \ion{Fe}{2} (``Iron Lo-BALs'').
These QSOs were discovered in optical follow-ups to a deep radio survey;
FIRST J155633.8+351758 is radio-loud, the first BAL QSO so identified.

FIRST J084044.5+363328 is highly polarized and exhibits many
features found in other BAL QSOs.  The continuum is $\approx$4\% polarized near 
2000 \AA\ rest-frame, falling to $\approx$2\% at longer wavelengths, 
at a position angle of $\approx$50\arcdeg.  
The emission lines are unpolarized.
The polarization rises to $\approx$8\% in the low-ionization 
troughs of \ion{Mg}{2} $\lambda$2800 and \ion{Al}{3} $\lambda$1860.
The polarization and its position angle vary in a complicated manner
across the metastable \ion{Fe}{2} absorption lines, suggesting that more than
one mechanism is at work, or that the system geometry is complex.

FIRST J155633.8+351758 may be the most highly polarized BAL QSO known, 
and exhibits other unusual polarization properties compared to other 
highly polarized BAL QSOs.  The continuum is $\approx$13\% 
polarized near 2000 \AA\ rest-frame, falling to $\approx$7\% at longer 
wavelengths, at a position angle of 153\arcdeg.
The emission lines are polarized like 
the continuum, but in the absorption troughs the polarization 
drops to zero.  Currently available data cannot yet discriminate among
the possible lines of sight to BAL QSOs (edge-on, pole-on, or random).

\end{abstract}
\vfil\eject
\section{Introduction}

Becker et al. (1997) reported the discovery of two unusual low-ionization 
broad absorption line (BAL) QSOs, FIRST J084044.5+363328 
(hereafter 0840+3633, at $z=1.238$) and FIRST J155633.8+351758 (hereafter 
1556+3517 at $z=1.497$)\footnote{We adopt these redshifts based on 
comparing our spectra with composite QSO spectra, but an unambiguous redshift is 
difficult to obtain.},
from programs to obtain optical spectra of radio-selected
QSO candidates from the VLA FIRST Survey (Becker et al. 1995).
Both BAL QSOs exhibit narrow absorption lines from metastable excited
levels of \ion{Fe}{2} and \ion{Fe}{3} like Q 0059$-$2735 
(Hazard et al. 1987), the prototype of this rare class (``iron Lo-BAL'' QSOs, 
as coined by Becker et al. 1997).
Furthermore, FIRST 1556+3517 qualifies as ``radio-loud,''
the first BAL QSO so identified. 

Egami et al. (1996) presented near-IR spectra of Q 0059$-$2735 and
Hawaii 167 (see also Cowie et al. 1995), a high-$z$ iron Lo-BAL QSO found 
in a deep $K$\ band survey.
Both are red (Q 0059$-$2735 has $B-K=3.4$, Hawaii 167 has $B-K=5.25$),
and have large Balmer decrements
(H$\alpha$/H$\beta$ = 7.6 and 13, respectively).  Egami et al. argued that
the implied reddening was so high that much of the rest-frame UV light
must arise from scattered light or from a starburst.  In Hawaii 167 there
is evidence for a 4000 \AA\ break and no UV emission lines are apparent,
supporting the latter claim.
Near-IR photometry shows that the FIRST BAL QSOs are also red 
(Hall et al. 1997): FIRST 0840+3633 only modestly so ($B-K=3.29$), 
similar to Q 0059$-$2735,
but FIRST 1556+3517 ($B-K=6.57$) is among the reddest QSOs known.  
Low-ionization BAL QSOs appear moderately reddened compared with 
high-ionization BAL QSOs (Sprayberry \& Foltz 1992), so perhaps it is not 
surprising that these highly absorbed iron Lo-BAL QSOs are red.

BAL QSOs are a highly polarized subclass of high luminosity AGN.
Hines \& Schmidt (1997) find that 9 of 28 BAL QSOs have
a broadband optical polarization of 2\% or greater compared to
only 2 of 115 radio-quiet non-BAL QSOs.
Spectropolarimetry has been published for about ten BAL QSOs, 
including Lo-BAL QSOs (e.g., Schmidt, Hines, \& Smith 1997; Ogle 1997; Hines \& 
Wills 1995; Cohen et al. 1995; Goodrich \& Miller 1995; Glenn et al. 1994).
The polarimetric properties are similar from object to object:
continuum polarization as high as 5\%, increasing toward shorter wavelengths,
less polarized or unpolarized emission lines, and absorption troughs with
polarizations greater than or equal to the continuum.
These results have often been interpreted in terms of orientation
(Hines \& Wills 1995; Cohen et al. 1995; Goodrich \& Miller 1995):
BAL QSOs are normal QSOs seen along a line of sight skimming the edge of an
obscuring torus, with BAL clouds accelerated from the surface of the torus by
a wind, and polarized continuum light scattered above the torus along
a less obscured path.  Lo-BAL QSOs are then QSOs seen along the most absorbed
and dusty lines of sight.

We report spectropolarimetry of the iron Lo-BALs FIRST 0840+3633 
and FIRST 1556+3517. 


\section{Observations and Results}

On U. T. 1996 December 10,
we observed FIRST 0840+3633 with the Low Resolution Imaging Spectrometer (Oke
et al. 1995) in spectropolarimetry mode (Good\-rich, Cohen, \& Putney 1995)
on the 10 meter Keck II telescope.  
We used a 300 line mm$^{-1}$ grating blazed at 5000 \AA,
that, with the 1 $\arcsec$ slit (at the parallactic angle),
gave an effective resolution of 10 \AA\ (FWHM of lamp lines);
the dispersion was 2.5 \AA\ pixel$^{-1}$.
The seeing was $<1^{\prime\prime}$.
The observation was broken into four 5 minute exposures, one for each
waveplate position (0$\arcdeg$, 45$\arcdeg$, 22.5$\arcdeg$, 67.5$\arcdeg$).
Although we observed our calibration standards with and without
an order-blocking filter, we did not so observe the QSO, and the red end of 
the spectrum is weakly contaminated by second order light.
We observed FIRST 1556+3517 on U. T. 1997 April 9 with the
same set-up and conditions, and a exposure time for each 
waveplate position of 15 minutes. 
We used standard reduction techniques inside the IRAF NOAO package and 
the procedures of Miller, Robinson, \& Goodrich (1988) for
calculating Stokes parameters and errors.  Table 1 and figures 1 and 2 give
the results. 

FIRST 0840+3633 is a highly polarized BAL QSO and shares many of the
characteristics of previously studied BAL QSOs (e.g., Glenn et al. 1994;
Goodrich \& Miller 1995; Cohen et al. 1995; Hines \& Wills 1995).
These include: a polarized continuum with the polarization
increasing toward shorter wavelengths (from 2\% to 4\%), unpolarized emission
lines (although \ion{C}{3}] $\lambda$1909 may be polarized), 
and increased polarization in the Lo-BAL troughs of \ion{Al}{3}
$\lambda$1860 and \ion{Mg}{2} $\lambda$2800 (up to 8\%). 
The continuum polarization position angle is about 50\arcdeg\, 
although some rotation may be present in the absorption troughs.
The polarization structure is complex across the blended narrow
absorption-line troughs that include lines of metastable \ion{Fe}{2} and other
species.  The polarization rises (only to $\approx$4\%) in the \ion{Fe}{2} 
$\lambda$2750* and \ion{Fe}{2} $\lambda$2600* troughs, consistent
with reduced source coverage compared with the Lo-BAL troughs (less diluting
light is absorbed).
The trough of \ion{Fe}{2} $\lambda$2380* shows a drop in polarization and 
a significant position angle rotation.  The weaker absorption features
do not show significant polarization changes relative to the continuum.

FIRST 1556+3517 is also highly polarized, but differs in its 
polarization characteristics from other BAL QSOs.
The only common behavior is the rise in polarization toward 
shorter wavelengths (from 7\% to 13\%), and even this is sharper
than common.  The level of polarization is remarkable, among
the highest seen for a BAL QSO continuum polarization.
The polarization of the emission lines (most clearly seen for the prominent
\ion{Fe}{2} blend at 2950 \AA\ rest-frame) is the same as the continuum; 
in other BAL QSOs, the emission lines are less polarized or unpolarized.
The polarization drops in the absorption line troughs, apparently to zero in
the low-ionization BALs and in the stronger blended \ion{Fe}{2} troughs.
In almost all other BAL QSOs, 
the trough polarization is the same as or greater than that of the continuum.

\section{Discussion}

\subsection{The Polarization}

Scattering by either dust or electrons probably produces the continuum 
polarization in both FIRST 1556+3517 and FIRST 0840+3633, as has been
suggested for other BAL QSOs.  
Both are at high Galactic latitudes 
and have polarization wavelength dependences inconsistent with polarization
by transmission through aligned grains in our galaxy or in their
host galaxy (e.g., Goodrich \& Cohen 1992 and references therein).
FIRST 1556+3517, while radio-loud and having a flat radio spectrum 
(Becker et al. 1997), shows polarization of its emission lines at the same
level and position angle as the continuum, ruling out synchrotron radiation 
as the origin for the polarization.

A rise in polarization toward shorter wavelengths, especially for wavelengths
near 2000 \AA\ rest-frame, is predicted by a number of dust scattering models
(e.g., Kartje 1995).  The rise may also result from the combination of 
wavelength-independent scattering by electrons or small dust grains,
plus dilution by some redder component.  In the case of FIRST 0840+3633,
the polarization shape probably represents dilution by unpolarized emission
from extensive \ion{Fe}{2} blends (Wills, Netzer, \& Wills
1985), as the polarization rises again at the red end of our spectrum 
(Table 1).

The deep absorption troughs of FIRST 1556+3517 are unpolarized, indicating
that the scattered light is completely absorbed; Markarian 231, which has
strong absorption (but which is not broad enough to qualify as a BAL under
the strict but somewhat arbitrary definition of Weymann et al. 1991), also
has a powerful starburst and shows a decrease in the polarization in
its absorption troughs (Smith et al. 1995).  The trough light of FIRST 1556+3517
is probably emitted by an underlying stellar component, or by 
direct reddened QSO light.  The shape of the unpolarized spectrum filling in the
trough bottoms, if interpolated between the troughs, is indeed red, but at too
low a level to account for the dramatic change in the polarization. 
The trough light is $\sim10\%$ of the flux near \ion{C}{3}] $\lambda$1909.
If essentially all the light near \ion{C}{3}] $\lambda$1909 is scattered
except for this unpolarized trough component (i.e., the scattering polarization
efficiency is 90\%), then approximately half the light at the
red end of our spectrum is scattered.  A directly viewed FIRST 0840+3633-shaped
QSO spectrum can account for this dilution if reddened by $A_V$ $\sim$ 2.5,
or an average QSO spectrum if reddened by $A_V$ $\sim$ 3
(with an SMC-type extinction and R = 3.1).
Dilution and wavelength-dependent dust scattering may both be present, 
but we cannot distinguish between these with current data.

Because the emission lines in FIRST 1556+3517 are polarized
identically to the continuum, we place the scattering medium well
outside the broad-line region (e.g., $\sim$ 1 parsec for this luminosity).   
This is in contrast to most BAL QSOs for which the scattering medium
must be coincident or smaller than the broad-line region (e.g., Cohen et al.
1995).

Resonance line scattering may contribute to the polarization, and may explain
the polarization of \ion{C}{3}] $\lambda$1909 (as in PHL 5200, 
Cohen et al. 1995), and the change in polarization and the position 
angle rotation in the troughs in FIRST 0840+3633.  
The rise in polarization in the FIRST 0840+3633 Lo-BAL troughs
may more simply indicate that the scattered continuum is less absorbed
than the unpolarized, diluting continuum.

Wampler et al. (1995) analyzed the individual broad and narrow absorption lines
in a high-resolution spectrum of Q 0059$-$2735 and concluded that there are 
low-ionization condensations in a hotter BAL flow which occult different parts
of the background emission regions.
Thus it is probably not surprising to see complex changes and rotations
in the metastable absorption troughs of FIRST 0840+3633.
There is no systematic relationship between excitation energy and 
trough polarization for the metastable Fe II troughs, so a simple 
geometry is not evident.  A high-resolution spectrum is required for 
further progress.

\subsection{The Geometry}

Given axisymmetry based on other classes of AGNs and the prospect of a radio
jet axis, BAL QSOs may be seen along three possible lines of sight:
\begin{enumerate}
\item{An {\em edge-on} ``torus-skimming'' line of sight.}
\item{A {\em pole-on} line of sight.}
\item{{\em Random} lines of sight.}
\end{enumerate}

A primary discriminant between geometries is that edge-on systems
are expected to have higher polarizations than pole-on systems. 
High-$z$ BAL QSOs are highly polarized as a class (Ogle 1997; Hines \& Schmidt 
1997) while high-$z$ non-BAL QSOs are not (Antonucci et al. 1996).
This provides statistical support for an edge-on geometry for BAL outflows.
The high polarization in the FIRST iron Lo-BALs QSOs also suggests an
edge-on geometry, and the polarization wavelength dependence resembles
that of edge-on dust scattering models (Kartje 1995).
It is therefore natural to propose that iron Lo-BAL QSOs fit into 
unified models as the most edge-on QSOs.

The jets of radio-loud QSOs permit another way to measure their orientation.
The relativistic beaming model for radio sources
(e.g., Orr \& Browne 1982) unifies core-dominated (flat spectrum)
and lobe-dominated (steep spectrum) radio sources by means of orientation:
core-dominant objects are those viewed close to the jet axis,
while lobe-dominant objects are those viewed at larger angles.
 
The combination of radio-selected BAL QSOs (for which a radio jet orientation
can be obtained) with high optical polarization (with a polarization position
angle) can eventually be used to test geometries.  The position angle rotation 
caused by resonance line scattering in 0226$-1024$ indicates that the 
scattering medium is distributed perpendicular to the BAL outflow (Ogle 1997).
While the relationship between the radio axis and the polarization axis is
model dependent, in lower luminosity AGNs, ``face-on'' Seyfert 1 galaxies
have optical polarizations with position angles parallel to their system axes,
while ``edge-on'' Seyfert 2 galaxies generally have polarization position 
angles perpendicular to their jet axes (e.g., Antonucci 1993).

FIRST 1556+3517 has a flat and variable radio spectrum and is unresolved in 
FIRST survey images (5 $\arcsec$), suggesting the radio axis is pointed close to
the line of sight.  This object may represent a relativistically 
boosted radio-quiet QSO (e.g., Falcke, Sherwood, \& Patnaik 1996), 
with BAL flow along the jet axis, a very different geometry from currently
popular models.  The radio spectra of radio-quiet BAL QSOs show a variety
of shapes, including both flat and steep spectra, with properties similar to
those of radio-quiet and radio-loud QSOs 
(Barvainis \& Lonsdale 1997), suggesting that BAL QSOs are randomly oriented.
Higher resolution maps to determine the position angle of the radio jet in
FIRST 1556+3517, if present, can clarify the situation in this individual QSO.
Statistics will still be needed to determine between random lines of sight
and a special line of sight.

The narrow metastable \ion{Fe}{2} absorption lines might originate in 
the narrow-line region (NLR), as suggested by Halpern et al. (1996) for the 
broad-lined radio galaxy Arp 102B.
If so, a more face-on view is required to place the NLR along the line of 
sight, assuming the biconical geometry common for nearby AGNs.
Narrow-line emission is typically very weak or absent from Lo-BAL QSOs
(Turnshek et al. 1997; Boroson \& Meyers 1992), but this could be
the result of geometric effects.  Netzer \& Laor (1993) proposed dusty NLR
clouds with anisotropic emission. 
If dusty NLR clouds are the metastable absorbers,
they could also account for the reddening. 

\section{Summary}

We have shown that FIRST 0840+3633 and FIRST 1556+3517 are highly
polarized BAL QSOs, with the continuum polarization rising steeply toward 
shorter wavelengths while keeping a constant position angle in the 
continuum.  Scattering is the likely polarization mechanism in both,
with the effects of some combination of dust and dilution leading to the
wavelength dependences seen. 
FIRST 0840+3633 shows unpolarized emission lines and increasing polarizations
in its BAL troughs,
but complex polarization
behavior across its narrow metastable troughs.  FIRST 1556+3517 shows the 
highest continuum polarization of any BAL QSO yet examined,
emission lines polarized the same as the continuum, and unpolarized 
absorption troughs.  
While our spectropolarimetry alone cannot discriminate among the possible lines
of sight to these BAL QSOs, the combination of a polarization position angle
with future high-resolution radio mapping may help determine their geometry.

\acknowledgments

We thank the Keck staff, Bob Becker and his FIRST Survey collaborators,
Bev Wills, and Dean Hines.
The W. M. Keck Observatory is a scientific partnership between the
University of California and the California Institute of Technology,
made possible by the generous gift of the W. M. Keck Foundation.
This work has been performed under the auspices of the U.S. Department of Energy
by Lawrence Livermore National Laboratory under Contract W-7405-ENG-48.


\clearpage
\figcaption{FIRST 0840+3633.  The top abscissa
shows rest-frame wavelengths, while the bottom abscissae show observed-frame
wavelengths in \AA.  The binning is 10 \AA.
The top panel is the total flux spectrum (in
ergs s$^{-1}$ cm$^{-2}$ \AA$^{-1}$), and several absorption and emission lines 
are labeled (see Hazard et al. 1987 and Wampler et al. 1995 for 
detailed identifications in Q 0059$-$2735). 
The second panel down shows the polarization (the positive
definite bias is negligible).
The third panel is the polarization position angle in degrees.
The bottom panel shows the polarized flux, the product of the top two panels.
Error bars are 1 $\sigma$.}

\figcaption{FIRST 1556+3517.  The axes are
the same as in Figure 1, except that the polarization given is the 
Stokes $Q^{\prime}$ ($Q$ in the rotated coordinate frame for which $Q$ is 
aligned with the continuum position angle), and the polarized flux is the 
Stokes flux: $Q^{\prime}$ multiplied by the total flux.
The binning is 10 \AA, and error bars are omitted.}


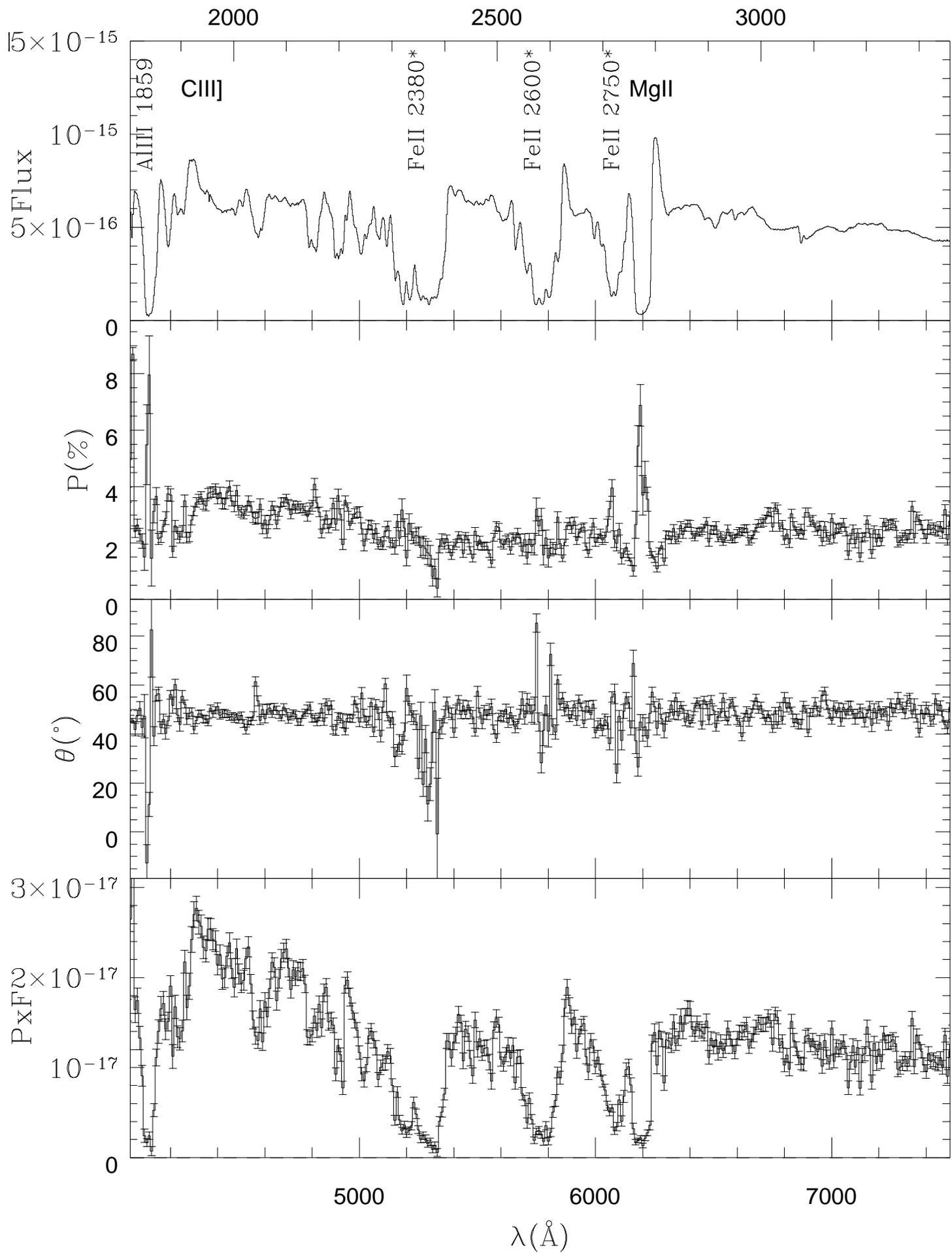
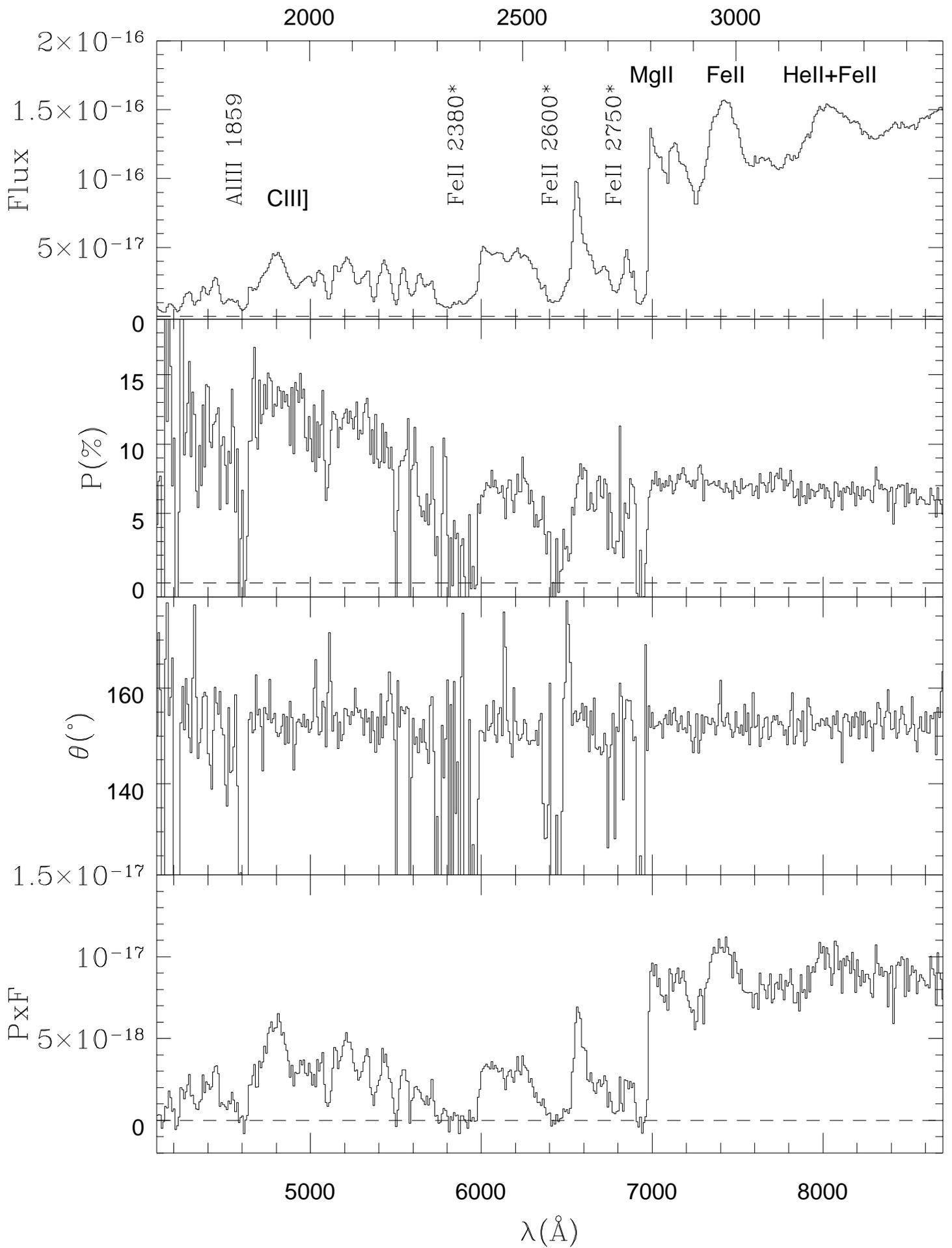
\end{document}